\documentclass[aps,prl,twocolumn,superscriptaddress,
  showpacs,preprintnumbers,amsmath,amssymb]{revtex4}
\usepackage{graphicx} 
\usepackage{dcolumn}  
\usepackage{rotating}
\usepackage{longtable}
\usepackage{amsmath}
\usepackage{multirow}
\usepackage{xcolor}
\usepackage[normalem]{ulem}
\RequirePackage{xspace}
\usepackage{relsize}
\graphicspath{{ps}}
\usepackage{enumerate}
\usepackage{hypernat}
\usepackage{hyperref}

\def\etal{{\it et al.}\xspace}
\def\cm   {\ensuremath{{\rm \,cm}}\xspace}
\def\invfb{\ensuremath{\mbox{\,fb}^{-1}}\xspace}
\def\Y#1S{\ensuremath{\Upsilon{(#1S)}}\xspace}
\newcommand{\gev}{\ensuremath{\mathrm{\,Ge\kern -0.1em V}}\xspace}
\newcommand{\mev}{\ensuremath{\mathrm{\,Me\kern -0.1em V}}\xspace}
\newcommand{\gevc}{\ensuremath{{\mathrm{\,Ge\kern -0.1em V\!/}c}}\xspace}
\newcommand{\mevc}{\ensuremath{{\mathrm{\,Me\kern -0.1em V\!/}c}}\xspace}
\newcommand{\gevcc}{\ensuremath{{\mathrm{\,Ge\kern -0.1em V\!/}c^2}}\xspace}
\newcommand{\mevcc}{\ensuremath{{\mathrm{\,Me\kern -0.1em V\!/}c^2}}\xspace}
\def\Mbc{\ensuremath{M^{}_{\rm bc}}\xspace}

\begin{document}


\preprint{\vbox{ \hbox{   }
    \hbox{Belle Preprint 2018-11}
    \hbox{KEK Preprint 2018-17}
}}

\title{ \quad\\[1.0cm] Search for the lepton-flavor-violating decay $B^{0}\to K^{\ast 0} \mu^{\pm} e^{\mp}$}

\noaffiliation
\affiliation{University of the Basque Country UPV/EHU, 48080 Bilbao}
\affiliation{Beihang University, Beijing 100191}
\affiliation{University of Bonn, 53115 Bonn}
\affiliation{Brookhaven National Laboratory, Upton, New York 11973}
\affiliation{Budker Institute of Nuclear Physics SB RAS, Novosibirsk 630090}
\affiliation{Faculty of Mathematics and Physics, Charles University, 121 16 Prague}
\affiliation{Chonnam National University, Kwangju 660-701}
\affiliation{University of Cincinnati, Cincinnati, Ohio 45221}
\affiliation{Deutsches Elektronen--Synchrotron, 22607 Hamburg}
\affiliation{University of Florida, Gainesville, Florida 32611}
\affiliation{Key Laboratory of Nuclear Physics and Ion-beam Application (MOE) and Institute of Modern Physics, Fudan University, Shanghai 200443}
\affiliation{Justus-Liebig-Universit\"at Gie\ss{}en, 35392 Gie\ss{}en}
\affiliation{Gifu University, Gifu 501-1193}
\affiliation{II. Physikalisches Institut, Georg-August-Universit\"at G\"ottingen, 37073 G\"ottingen}
\affiliation{SOKENDAI (The Graduate University for Advanced Studies), Hayama 240-0193}
\affiliation{Gyeongsang National University, Chinju 660-701}
\affiliation{Hanyang University, Seoul 133-791}
\affiliation{University of Hawaii, Honolulu, Hawaii 96822}
\affiliation{High Energy Accelerator Research Organization (KEK), Tsukuba 305-0801}
\affiliation{J-PARC Branch, KEK Theory Center, High Energy Accelerator Research Organization (KEK), Tsukuba 305-0801}
\affiliation{Forschungszentrum J\"{u}lich, 52425 J\"{u}lich}
\affiliation{IKERBASQUE, Basque Foundation for Science, 48013 Bilbao}
\affiliation{Indian Institute of Science Education and Research Mohali, SAS Nagar, 140306}
\affiliation{Indian Institute of Technology Bhubaneswar, Satya Nagar 751007}
\affiliation{Indian Institute of Technology Guwahati, Assam 781039}
\affiliation{Indian Institute of Technology Hyderabad, Telangana 502285}
\affiliation{Indian Institute of Technology Madras, Chennai 600036}
\affiliation{Indiana University, Bloomington, Indiana 47408}
\affiliation{Institute of High Energy Physics, Chinese Academy of Sciences, Beijing 100049}
\affiliation{Institute of High Energy Physics, Vienna 1050}
\affiliation{Institute for High Energy Physics, Protvino 142281}
\affiliation{INFN - Sezione di Napoli, 80126 Napoli}
\affiliation{INFN - Sezione di Torino, 10125 Torino}
\affiliation{Advanced Science Research Center, Japan Atomic Energy Agency, Naka 319-1195}
\affiliation{J. Stefan Institute, 1000 Ljubljana}
\affiliation{Institut f\"ur Experimentelle Teilchenphysik, Karlsruher Institut f\"ur Technologie, 76131 Karlsruhe}
\affiliation{Kennesaw State University, Kennesaw, Georgia 30144}
\affiliation{King Abdulaziz City for Science and Technology, Riyadh 11442}
\affiliation{Department of Physics, Faculty of Science, King Abdulaziz University, Jeddah 21589}
\affiliation{Korea Institute of Science and Technology Information, Daejeon 305-806}
\affiliation{Korea University, Seoul 136-713}
\affiliation{Kyoto University, Kyoto 606-8502}
\affiliation{Kyungpook National University, Daegu 702-701}
\affiliation{LAL, Univ. Paris-Sud, CNRS/IN2P3, Universit\'{e} Paris-Saclay, Orsay}
\affiliation{\'Ecole Polytechnique F\'ed\'erale de Lausanne (EPFL), Lausanne 1015}
\affiliation{P.N. Lebedev Physical Institute of the Russian Academy of Sciences, Moscow 119991}
\affiliation{Faculty of Mathematics and Physics, University of Ljubljana, 1000 Ljubljana}
\affiliation{Ludwig Maximilians University, 80539 Munich}
\affiliation{Luther College, Decorah, Iowa 52101}
\affiliation{University of Malaya, 50603 Kuala Lumpur}
\affiliation{University of Maribor, 2000 Maribor}
\affiliation{Max-Planck-Institut f\"ur Physik, 80805 M\"unchen}
\affiliation{School of Physics, University of Melbourne, Victoria 3010}
\affiliation{University of Mississippi, University, Mississippi 38677}
\affiliation{University of Miyazaki, Miyazaki 889-2192}
\affiliation{Moscow Physical Engineering Institute, Moscow 115409}
\affiliation{Moscow Institute of Physics and Technology, Moscow Region 141700}
\affiliation{Graduate School of Science, Nagoya University, Nagoya 464-8602}
\affiliation{Kobayashi-Maskawa Institute, Nagoya University, Nagoya 464-8602}
\affiliation{Universit\`{a} di Napoli Federico II, 80055 Napoli}
\affiliation{Nara Women's University, Nara 630-8506}
\affiliation{National Central University, Chung-li 32054}
\affiliation{National United University, Miao Li 36003}
\affiliation{Department of Physics, National Taiwan University, Taipei 10617}
\affiliation{H. Niewodniczanski Institute of Nuclear Physics, Krakow 31-342}
\affiliation{Nippon Dental University, Niigata 951-8580}
\affiliation{Niigata University, Niigata 950-2181}
\affiliation{Novosibirsk State University, Novosibirsk 630090}
\affiliation{Osaka City University, Osaka 558-8585}
\affiliation{Pacific Northwest National Laboratory, Richland, Washington 99352}
\affiliation{Panjab University, Chandigarh 160014}
\affiliation{Peking University, Beijing 100871}
\affiliation{University of Pittsburgh, Pittsburgh, Pennsylvania 15260}
\affiliation{Punjab Agricultural University, Ludhiana 141004}
\affiliation{Theoretical Research Division, Nishina Center, RIKEN, Saitama 351-0198}
\affiliation{University of Science and Technology of China, Hefei 230026}
\affiliation{Showa Pharmaceutical University, Tokyo 194-8543}
\affiliation{Soongsil University, Seoul 156-743}
\affiliation{University of South Carolina, Columbia, South Carolina 29208}
\affiliation{Stefan Meyer Institute for Subatomic Physics, Vienna 1090}
\affiliation{Sungkyunkwan University, Suwon 440-746}
\affiliation{School of Physics, University of Sydney, New South Wales 2006}
\affiliation{Department of Physics, Faculty of Science, University of Tabuk, Tabuk 71451}
\affiliation{Tata Institute of Fundamental Research, Mumbai 400005}
\affiliation{Department of Physics, Technische Universit\"at M\"unchen, 85748 Garching}
\affiliation{Toho University, Funabashi 274-8510}
\affiliation{Department of Physics, Tohoku University, Sendai 980-8578}
\affiliation{Earthquake Research Institute, University of Tokyo, Tokyo 113-0032}
\affiliation{Department of Physics, University of Tokyo, Tokyo 113-0033}
\affiliation{Tokyo Institute of Technology, Tokyo 152-8550}
\affiliation{Tokyo Metropolitan University, Tokyo 192-0397}
\affiliation{Virginia Polytechnic Institute and State University, Blacksburg, Virginia 24061}
\affiliation{Wayne State University, Detroit, Michigan 48202}
\affiliation{Yamagata University, Yamagata 990-8560}
\affiliation{Yonsei University, Seoul 120-749}
  \author{S.~Sandilya}\affiliation{University of Cincinnati, Cincinnati, Ohio 45221} 
  \author{K.~Trabelsi}\affiliation{High Energy Accelerator Research Organization (KEK), Tsukuba 305-0801}\affiliation{SOKENDAI (The Graduate University for Advanced Studies), Hayama 240-0193} 
  \author{A.~J.~Schwartz}\affiliation{University of Cincinnati, Cincinnati, Ohio 45221} 
  \author{I.~Adachi}\affiliation{High Energy Accelerator Research Organization (KEK), Tsukuba 305-0801}\affiliation{SOKENDAI (The Graduate University for Advanced Studies), Hayama 240-0193} 
  \author{H.~Aihara}\affiliation{Department of Physics, University of Tokyo, Tokyo 113-0033} 
  \author{S.~Al~Said}\affiliation{Department of Physics, Faculty of Science, University of Tabuk, Tabuk 71451}\affiliation{Department of Physics, Faculty of Science, King Abdulaziz University, Jeddah 21589} 
  \author{D.~M.~Asner}\affiliation{Brookhaven National Laboratory, Upton, New York 11973} 
  \author{H.~Atmacan}\affiliation{University of South Carolina, Columbia, South Carolina 29208} 
  \author{V.~Aulchenko}\affiliation{Budker Institute of Nuclear Physics SB RAS, Novosibirsk 630090}\affiliation{Novosibirsk State University, Novosibirsk 630090} 
  \author{T.~Aushev}\affiliation{Moscow Institute of Physics and Technology, Moscow Region 141700} 
% ...
\collaboration{Belle Collaboration}

\noaffiliation

\begin{abstract}
We have searched for the lepton-flavor-violating decay $B^{0}\to K^{\ast 0} \mu^{\pm} e^{\mp}$ using a data sample of 711~\invfb that contains $772 \times 10^{6}$ $B\bar{B}$ pairs. The data were collected near the \Y4S resonance with the Belle detector at the KEKB asymmetric-energy $e^{+}e^{-}$ collider. No signals were observed, and we set 90\% confidence level upper limits on the branching fractions of ${\cal B}(B^{0}\to K^{\ast 0} \mu^{+} e^{-})< 1.2\times 10^{-7}$, ${\cal B}(B^{0}\to K^{\ast 0} \mu^{-} e^{+})< 1.6\times 10^{-7}$, and, for both decays combined, ${\cal B}(B^{0}\to K^{\ast 0} \mu^{\pm} e^{\mp}) < 1.8\times 10^{-7}$. These are the most stringent limits on these decays to date.
\end{abstract}


\maketitle

{\renewcommand{\thefootnote}{\fnsymbol{footnote}}}
\setcounter{footnote}{0}

In recent years, measurements from the LHCb~\cite{{btosll:lhcb},{btosll:lhcb1}} experiment have exhibited possible deviations from lepton universality in flavor-changing neutral-current $b \to s \ell^+\ell^-$ transitions. Such universality is an important symmetry of the Standard Model. These deviations have generated much interest within the theoretical community, and several models of new physics~\cite{{btosll:th1},{btosll:th2},{btosll:th3},{btosll:th4},{btosll:th5},{btosll:th6},{btosll:th7},{btosll:th8}} have been proposed to explain these discrepancies. In many such models, violation of lepton universality is accompanied by lepton flavor violation (LFV)~\cite{lfv:th1}. The idea of LFV in $B$ decays was discussed in Refs.~\cite{{lfv:th2},{lfv:th3},{lfv:th4},{lfv:th5},{lfv:th6},{lfv:th7},{lfv:th8},{lfv:th9}}. Experimentally, one way to search for LFV is via the decays $B^{0}\to K^{\ast 0} \mu^{\pm} e^{\mp}$~\cite{kst892}, which have large available phase space and also avoid the helicity suppression that a two-body decay such as $B^{0}\to\mu^{\pm} e^{\mp}$ might be subjected to. The most stringent upper limits for $B^{0}\to K^{\ast 0} \mu^{\pm} e^{\mp}$ were set by the BaBar experiment based on a data sample of $229 \times 10^{6}$ $B\bar{B}$ events~\cite{lfv:babar}. Here, we report a search for $B^{0}\to K^{\ast 0} \mu^{\pm} e^{\mp}$ using a data sample of $(772\pm 11) \times 10^6$ $B\bar{B}$ events (711 \invfb), which is more than 3 times larger than that of BaBar. The data sample was collected by the Belle experiment running near the \Y4S resonance at the KEKB $e^{+}e^{-}$ collider~\cite{KEKB}.

The Belle detector is a large-solid-angle magnetic spectrometer consisting of a silicon vertex detector (SVD), a 50-layer central drift chamber (CDC), an array of aerogel threshold Cherenkov counters (ACC), a barrel-like arrangement of time-of-flight scintillation counters (TOF), and an electromagnetic calorimeter (ECL) comprising CsI(Tl) crystals. All are located inside a superconducting solenoid coil which provides a 1.5 T magnetic field. An iron flux return yoke located outside the coil is instrumented with resistive-plate chambers (KLM) to detect $K^{0}_{L}$ mesons and muons. Further details of the detector are given in Ref.~\cite{belle:detector}. Two inner detector configurations were used: a 2.0 \cm radius beam-pipe and a three-layer SVD were used to record the first sample of 140\invfb , while a 1.5 \cm radius beam-pipe, a four-layer SVD, and a small-cell inner drift chamber were used to record the remaining 571\invfb~\cite{nbb}.

To study properties of signal events and optimize selection criteria, we generate samples of Monte Carlo (MC) simulated events. These samples are generated with the \textsc{EvtGen} package~\cite{evtgen} using three-body phase space and assuming that the $K^{\ast 0}$ is unpolarized. The detector response is simulated with the \textsc{GEANT3} package~\cite{geant3}.

We begin reconstructing $B^{0}\to K^{\ast 0} \mu^{\pm} e^{\mp}$~\cite{cc} decays by selecting charged particles that originate from a region near the $e^+e^-$ interaction point. This region is defined using impact parameters: we require $dr < 1$~cm in the $x$-$y$ plane (transverse to the positron beam), and $|dz| < 4\cm$ along the $z$ axis (antiparallel to the positron beam). To reduce backgrounds from low-momentum particles, we require that tracks have a transverse momentum ($p^{}_T$) greater than~0.1~$\gevc$. 

From selected tracks, we identify $K^{\pm}$, $\pi^{\pm}$, $\mu^{\pm}$, and $e^{\pm}$ candidates using information from the CDC, ACC, and TOF detectors. The $K^{\pm}$ and $\pi^{\pm}$ candidates are identified by constructing the likelihood ratio ${\cal R}^{}_{K} = {\cal L}^{}_K / ({\cal L}^{}_K + {\cal L}^{}_\pi )$, where ${\cal L}^{}_{\pi}$ and ${\cal L}^{}_{K}$ are relative likelihoods for kaons and pions, respectively, calculated based on the number of photoelectrons in the ACC, the specific ionization in the CDC, and the time-of-flight as determined from TOF hit times. We select kaons (pions) by requiring ${\cal R}^{}_{K} > 0.6$  ($< 0.4$). This criterion is 92\% (89\%) efficient for kaons (pions), and has a misidentification rate of 7\% (8\%) for pions (kaons).

Muon candidates are identified based on information from the KLM detector. We require that candidates have momentum greater than 0.8~\gevc, and that they have a penetration depth and degree of transverse scattering consistent with a muon, given the track momentum measured in the CDC~\cite{muid}. A criterion on normalized muon likelihood, ${\cal R}_{\mu} > 0.9$, is used to select muon candidates. For this requirement the average muon detection efficiency is 89\%, and the average pion misidentification rate is 1.4\%~\cite{pid}.

Electron candidates are required to have momentum greater than 0.4~\gevc and are identified using the following information: the ratio of ECL energy to the CDC track momentum; the ECL shower shape; position matching between the CDC track and the ECL cluster; the energy loss in the CDC; and the response of the ACC~\cite{eid}. A requirement on normalized electron likelihood ${\cal R}^{}_{e} >~0.9$ is imposed. This requirement has an efficiency of 92\% and a pion mis-identification rate of about 0.25\%~\cite{pid}. To recover electron energy lost due to possible bremsstrahlung, we search for photons inside a cone of radius $50~\rm mrad$ centered around the electron momentum. If a photon is found within this cone, its four-momentum is added to that of the electron.

Kaon and pion candidates are combined to form $K^{\ast 0}$ candidates by requiring that their $K$-$\pi$ invariant mass be within a $100\mevcc$ window centered around the $K^{\ast 0}$ mass~\cite{pdg}. $B$~candidates are subsequently reconstructed by combining $K^{\ast 0}$, $\mu^\pm$, and $e^\mp$ candidates. To discriminate signal decays from background, two kinematic variables are defined: the beam-energy-constrained mass $\Mbc = \sqrt{(E^{}_{\rm beam}/c^{2})^{2} - (p^{}_{B}/c)^{2}}$, and the energy difference $\Delta E =  E^{}_{B} - E^{}_{\rm beam}$, where $E^{}_{\rm beam}$ is the beam energy and  $E^{}_{B}$ and $p^{}_{B}$ are the energy and momentum, respectively, of the $B$ candidate. All of these quantities are evaluated in the $e^{+}e^{-}$ center-of-mass (CM) frame. For signal events, the $\Delta E$ distribution peaks near zero, and the \Mbc distribution peaks near the $B$ mass. We retain events satisfying the loose requirements $\Delta E \in [-0.05, 0.04] \gev$ and $\Mbc > 5.2~\gevcc$.

After the above selection criteria are imposed, about 3\% of events have more than one signal $B$ candidate. To select a single candidate, we choose the one with the smallest $\chi^{2}$ from a vertex fit of the four charged tracks. From MC simulation, we find that this criterion identifies the correct signal decay 63\% of the time.

At this stage of the analysis, there is significant background from $e^+e^-\to q\bar{q}$ $(q=u,d,s,c)$ continuum events. As lighter quarks are produced with large initial momentum, these events tend to consist of two back-to-back jets of pions and kaons. In contrast, $e^+e^-\to b\bar{b}$ events result in $B\bar{B}$ pairs produced almost at rest in the CM frame; this results in more spherically distributed daughter particles. We thus distinguish $B\bar{B}$ events from $q\bar{q}$ background based on event topology. We use a multivariate analyzer constructed from a neural network (NN) that uses the following information:

\begin{enumerate}[ {(}1{)} ]

\item A likelihood ratio constructed from modified Fox-Wolfram moments~\cite{{KSFW},{FW}}.

\item The angle between the thrust axis of the $B$ decay products and that of the rest of the event (the thrust axis being defined as the direction that maximizes the sum of the longitudinal momenta of all particles). 

\item The angle $\theta^{}_B$ between the $z$ axis and the $B$ flight direction in the CM frame (for $B\bar{B}$ events, $dN/d\cos\theta^{}_B\propto 1-\cos^{2}\theta_{B}$, whereas for continuum events, $dN/d\cos\theta^{}_B\approx {\rm\ constant}$).

\item Flavor-tagging information from the other (nonsignal) $B$ decay. Our flavor-tagging algorithm~\cite{belle:qr} outputs two variables: the flavor $q$ of the tag-side $B$, and the tag quality~$r$. The latter ranges from zero for no flavor information to one for unambiguous flavor assignment.

\end{enumerate}

We choose a selection criterion on the NN output (${\cal O}^{q\bar{q}}_{\rm NN}$) by optimizing a figure of merit $\varepsilon/\sqrt{N^{}_B}$, where $\varepsilon$ is the signal efficiency as determined from MC simulation, and $N^{}_B$ is the total number of background events expected in a restrictive signal region $\Mbc>5.27 \gevcc$. We obtain a criterion ${\cal O}^{q\bar{q}}_{\rm NN} > 0.5$, which rejects 94\% of $q\bar{q}$ background while retaining 73\% of signal events.

After this criterion is applied, the remaining background arises mainly from $B$ decays that produce two leptons. Such background falls into three categories: ({\it a})\ both $B$ and $\bar{B}$ decay semileptonically; ({\it b})\ a $B\to \bar{D}^{(*)}X\ell^+\nu$ decay is followed by a $\bar{D}^{(*)}\to X\ell^-\bar{\nu}$ decay; and ({\it c})\ hadronic $B$ decays where one or more daughter particles are misidentified as leptons. To suppress these backgrounds, we use a second NN that utilizes the following information:

\begin{enumerate}[ {(}1{)} ]

\item The separation in $z$ between the signal $B$ decay vertex and the vertex of the other $B$.
  
\item The sum of the ECL energy of tracks and clusters {\it not\/} associated with the signal $B$ decay.

\item The $\chi^{2}$ of the vertex fit of the four charged tracks forming the signal-$B$ decay vertex.
  
\item The separation in $z$ between the two lepton tracks. 

\end{enumerate}

The criterion on the NN output, ${\cal O}^{BB}_{\rm NN} > 0$, is obtained by maximizing the above figure of merit, $\varepsilon/\sqrt{N^{}_B}$. At this stage, we also optimize the criterion on the variable $\Delta E$, obtaining $|\Delta E| < 0.025 \gev$.

After applying this NN selection, only a small amount of background survives. We study this remaining background using MC simulation and find that the main source is $B^{0}\to K^{\ast 0} (\to\!K^+\pi^-)\,J/\psi (\to\!\ell^{+}\ell^{-})$ decays in which one of the leptons is misidentified and swapped with the $K^{+}$ or $\pi^{-}$. To suppress this background, we apply a set of vetoes. For $B^{0}\to K^{\ast 0} \mu^{+} e^{-}$ signal events, we apply three: the dilepton invariant mass must satisfy $M(\ell^{+}\ell^{-})\!\notin\![3.04,3.12]$\gevcc; the kaon-electron invariant mass must satisfy $M(K^{+} e^{-})\!\notin\![2.90,3.12]$\gevcc; and the pion-muon invariant mass must satisfy $M(\pi^{-}\mu^{+})\!\notin\![3.06,3.12]$\gevcc. For $B^{0}\to K^{\ast 0} \mu^{-} e^{+}$ signal events, we apply two vetoes: the dilepton invariant mass must satisfy $M(\ell^{+}\ell^{-})\!\notin\![3.02,3.12]$\gevcc, and the pion-electron invariant mass must satisfy $M(\pi^{-} e^{+})\!\notin\![3.02,3.12]$\gevcc. While calculating these invariant masses, the mass hypothesis for a hadron is taken to be that of the associated lepton. These vetoes have relative efficiencies of 90.4\% and 94.8\% for $B^{0}\to K^{\ast 0} \mu^{+} e^{-}$ and $B^{0}\to K^{\ast 0} \mu^{-} e^{+}$, respectively. We use a high-statistics MC sample to study backgrounds originating from charmless hadronic $B$ decays and find them to be negligible. The largest contribution is from $B^{0}\to K^{\ast 0} \pi^{+} \pi^{-}$ in which the pions are misidentified as leptons; this contribution is only 0.01 event. To avoid bias, all selection criteria are determined in a ``blind'' manner, {\it i.e.,} they are finalized before looking at events in the signal region.

\begin{figure}[htbp]
  \mbox{\includegraphics[width=\columnwidth]{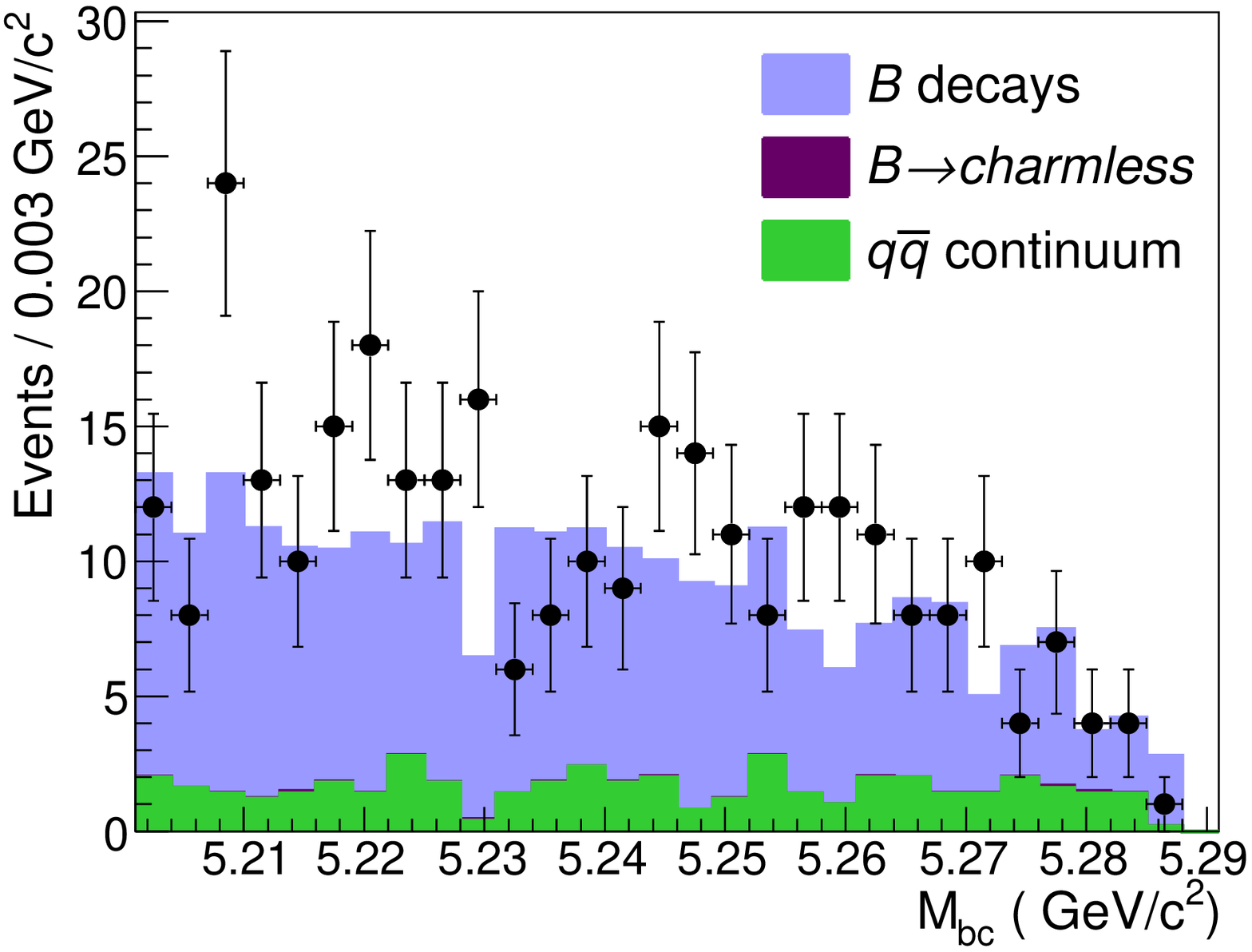}}
  \mbox{\includegraphics[width=\columnwidth]{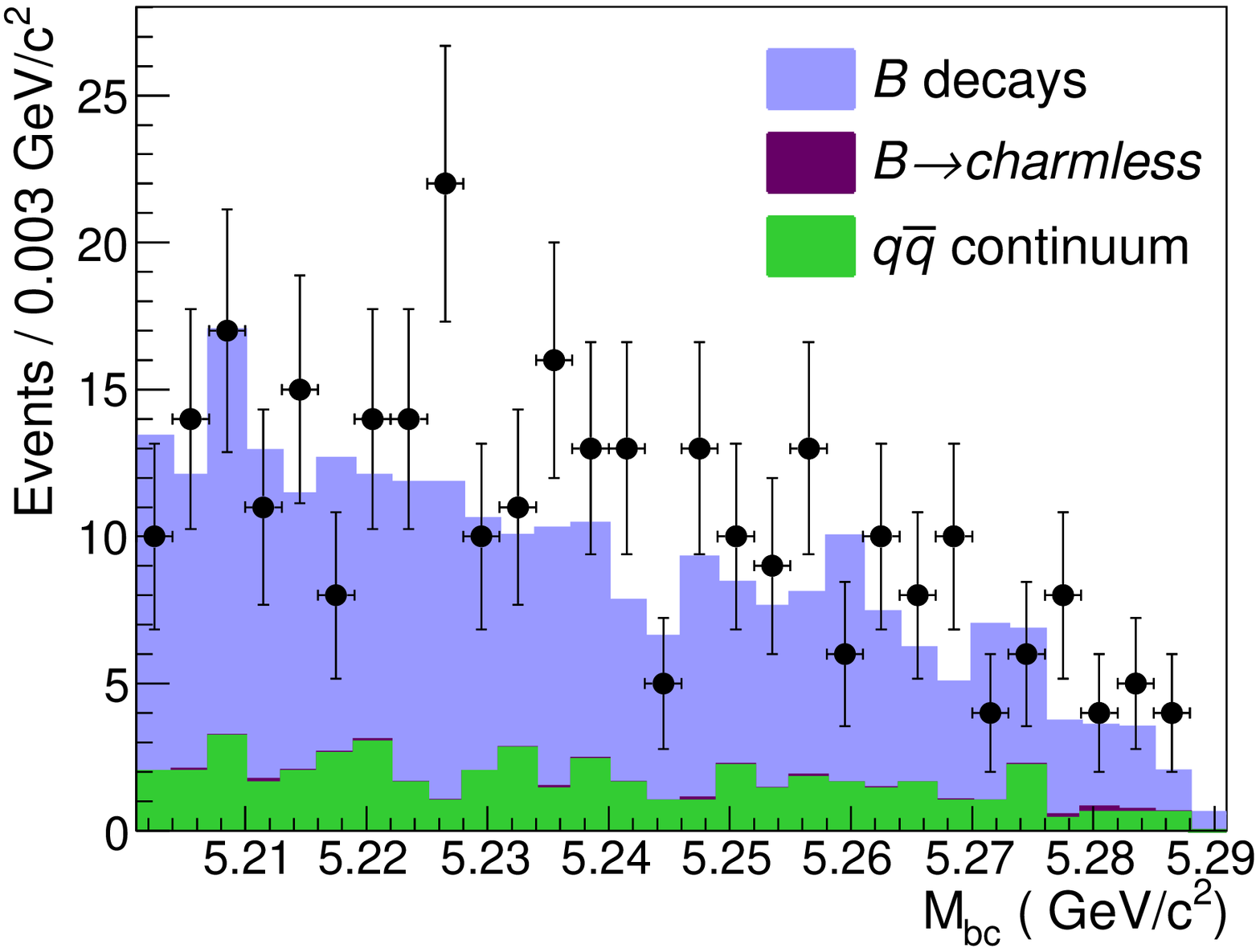}}
  \mbox{\includegraphics[width=\columnwidth]{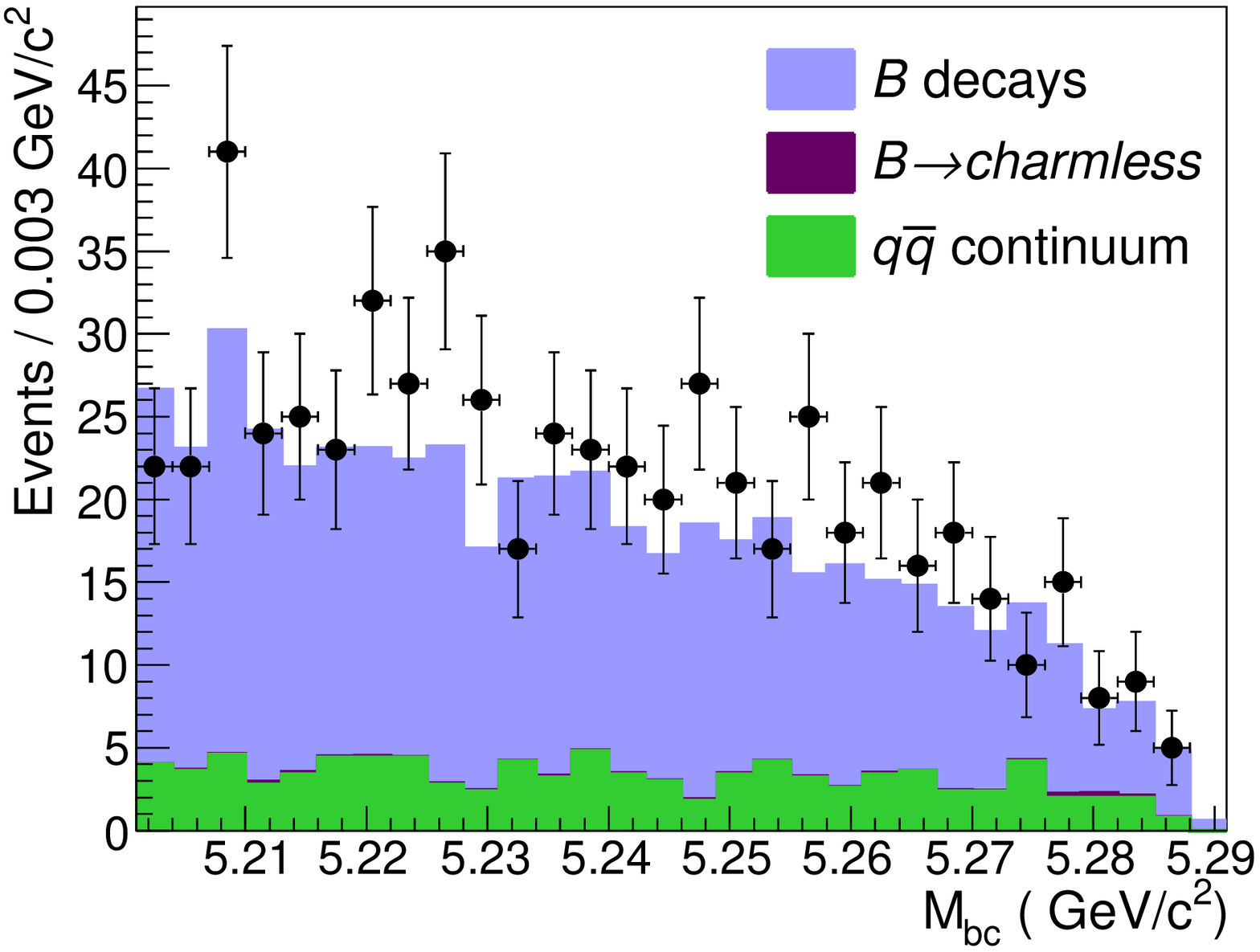}}
  
  \caption{The \Mbc distribution for data and MC events that pass the selection criteria for the decays $B^{0}\to K^{\ast 0} \mu^{+} e^{-}$ (top), $B^{0}\to K^{\ast 0} \mu^{-} e^{+}$ (middle), and for both decays combined (bottom). Points with error bars are the data, while the color filled stacked histograms depict MC components from generic $B$ decays (blue), $q\bar{q}$ continuum (green), and negligible contributions from charmless hadronic $B$ decays (purple).}
  \label{fig:data_mc}
\end{figure}

To test our understanding of remaining backgrounds, we compare the \Mbc distributions for data and MC events, as shown in Fig.~\ref{fig:data_mc}. The plots show good agreement between data and MC for both the number of events observed and the shapes of the distributions.

\begin{figure}[htbp]
  \mbox{\includegraphics[width=\columnwidth]{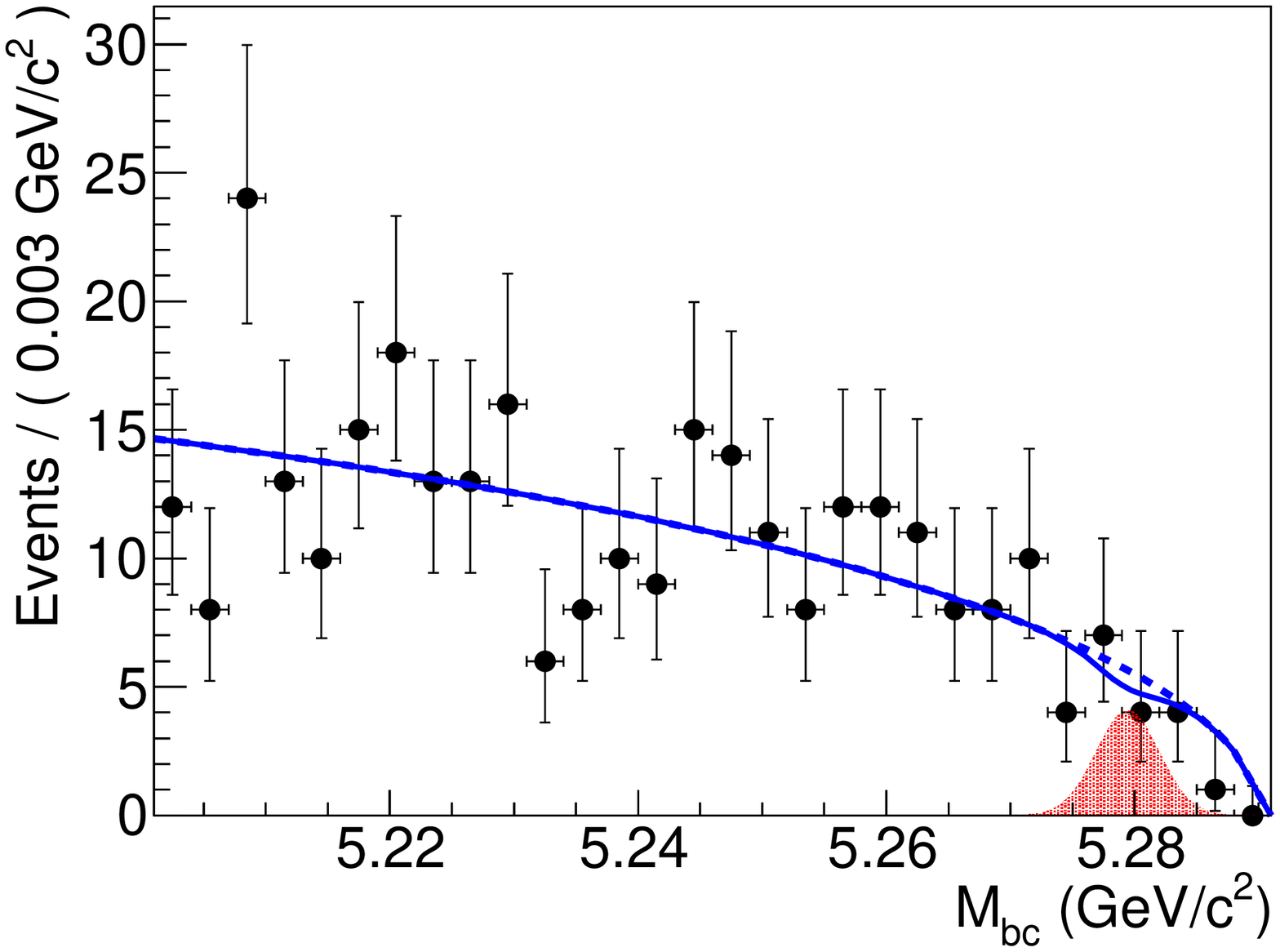}}
  \mbox{\includegraphics[width=\columnwidth]{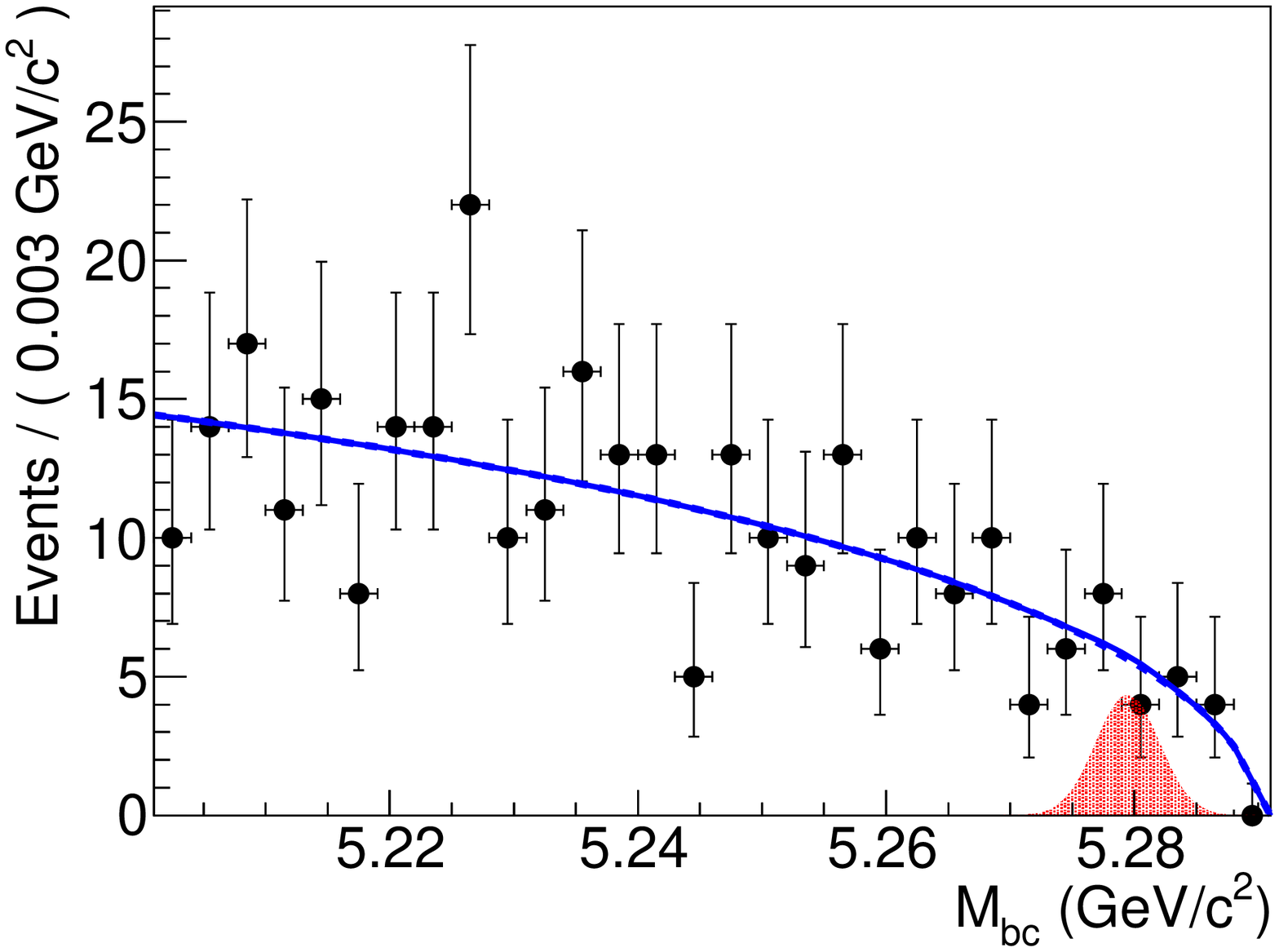}}
  \mbox{\includegraphics[width=\columnwidth]{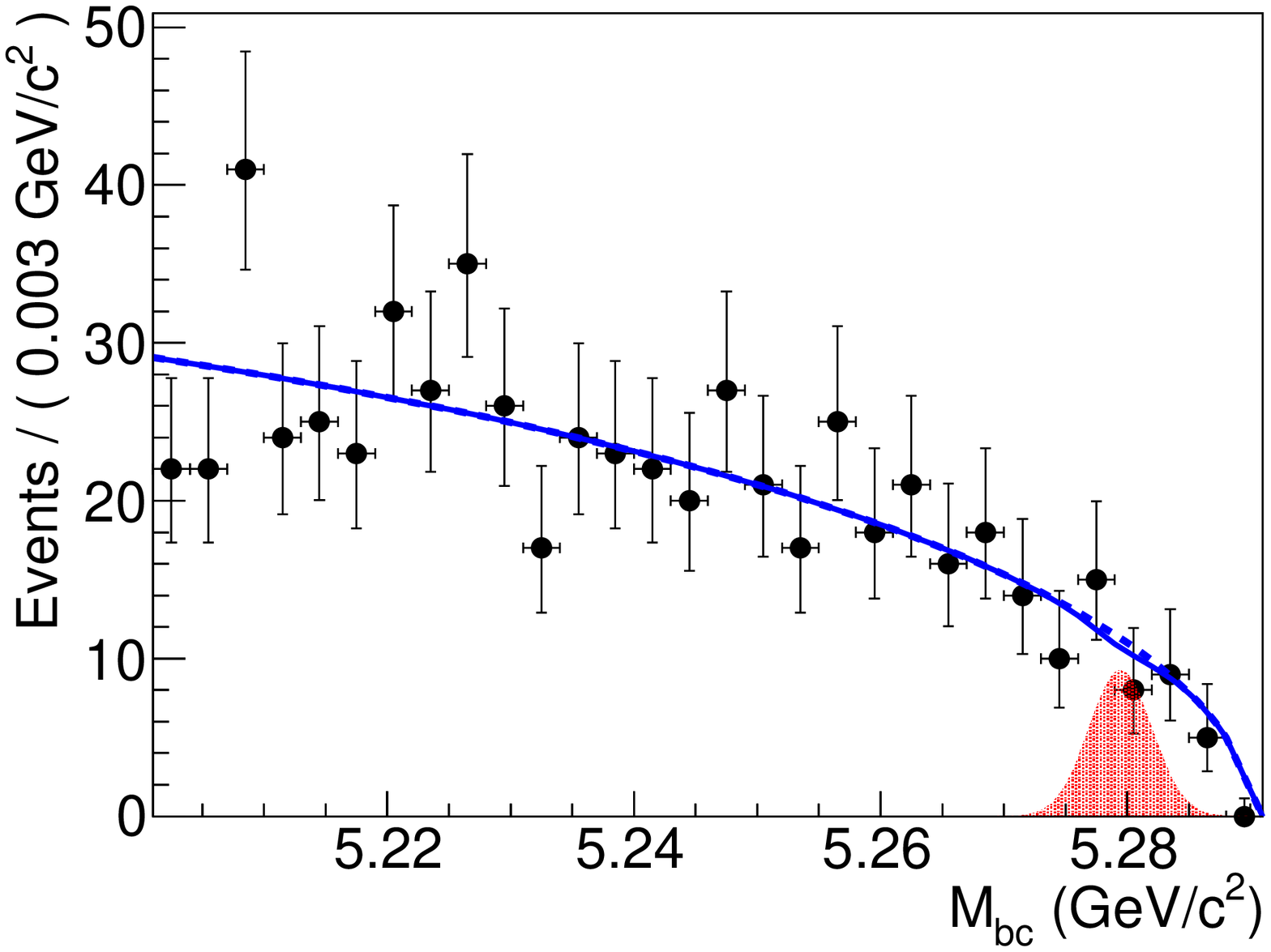}}
  
  \caption{The \Mbc distribution for data events that pass the selection criteria for the decays $B^{0}\to K^{\ast 0} \mu^{+} e^{-}$ (top), $B^{0}\to K^{\ast 0} \mu^{-} e^{+}$ (middle), and also both decays combined (bottom). Points with error bars are the data, and the blue solid curve is the result of the fit for the signal-plus-background hypothesis, where the blue dashed curve is the background component. The red shaded histogram represents the signal PDF with arbitrary normalization.}
\label{fig:mbc_plot}
\end{figure}

We calculate the signal yield by performing an unbinned extended maximum-likelihood fit to the \Mbc distribution. The probability density function (PDF) used to model signal decays is a Gaussian, and that for all backgrounds combined is an ARGUS function~\cite{argus}. The signal shape parameters are obtained from MC simulation. We check these parameters by fitting the \Mbc distribution of a control sample of $B^{0}\to K^{\ast 0} (\to\!K^+\pi^-)\,J/\psi (\to\!\ell^{+}\ell^{-})$ decays. For this control sample, we fit both data and MC events and find excellent agreement between them for the shape parameters obtained. All background shape parameters, along with the signal and background yields, are floated in the fit. The fitted \Mbc distributions are shown in Fig.~\ref{fig:mbc_plot}. The fitted yields are $N_{\rm sig} = -1.5\,^{+4.7}_{-4.1}$ and $0.4\,^{+4.8}_{-4.5}$ for $B^{0}\to K^{\ast 0} \mu^{+} e^{-}$ and $B^{0}\to K^{\ast 0} \mu^{-} e^{+}$, respectively. By combining both final states, we obtain $N_{\rm sig} = -1.2\,^{+6.8}_{-6.2}$. 

As there is no evidence of a signal, we calculate 90\% confidence level (C.L.) upper limits on the branching fractions using a frequentist method as follows. We scan through a range of possible signal yields, and for each yield generate 10 000 sets of signal and background events according to their PDFs. Each set of events is statistically equivalent to our data set of 711~fb$^{-1}$. We combine signal and background samples and perform our fitting procedure on these combined sets of events. We then calculate, for each input value of signal yield, the fraction of sets ($f^{}_{\rm sig}$) that have a fitted yield less than that observed in the data. The input signal having $f^{}_{\rm sig}=0.10$ is taken as an upper limit $N^{\rm UL}_{\rm sig}$ (statistical error only). We convert $N^{\rm UL}_{\rm sig}$ into an upper limit on the branching fraction (${\cal B}^{\rm UL}$) via the formula
\begin{eqnarray*}
{\cal B} & = & \frac{N_{\rm sig}}
{{\cal B}(K^{\ast 0} \to K^{+} \pi^{-})
\times 2\times N_{B\bar{B}} \times f^{00} \times \varepsilon}\,,
  \label{eq:bf}
\end{eqnarray*}
where ${\cal B}(K^{\ast 0} \to K^{+} \pi^{-}) = 0.6651$ is the assumed branching fraction (from isospin symmetry) for the intermediate decay $K^{\ast 0} \to K^{+} \pi^{-}$; $N_{B\bar{B}}$ is the number of $B\bar{B}$ pairs, $(7.72 \pm 0.11) \times 10^{8}$; $f^{00}$ is the branching fraction ${\cal B}(\Y4S \to B^{0}\bar{B^{0}}) = 0.486 \pm 0.006$~\cite{pdg}; and $\varepsilon$ is the signal reconstruction efficiency as calculated from MC simulation. We include systematic uncertainty in ${\cal B}^{\rm UL}$ by smearing the $N^{}_{\rm sig}$ distributions of the aforementioned statistically equivalent samples by the total fractional systematic uncertainty (see below) before calculating $f^{}_{\rm sig}$. The resulting upper limits are listed in Table~\ref{tab:ul-est}. For the upper limit on both decays $K^{\ast 0}\mu^+ e^-$ and $K^{\ast 0}\mu^- e^+$ combined, ${\cal B}(B^0\to K^{\ast 0}\mu^\pm e^\mp) \equiv {\cal B}(B^0\to K^{\ast 0}\mu^+ e^-) + {\cal B}(B^0\to K^{\ast 0}\mu^- e^+)$, and the branching fractions for the two modes are assumed to be identical when calculating the efficiency.

\begin{table}[tbh]
  \caption{Results from the fits. The rightmost columns correspond to efficiency, signal yield, 90\% C.L. upper limit on the signal yield, and 90\% C.L. upper limit on the branching fraction.}
  \label{tab:ul-est}
  \renewcommand{\arraystretch}{1.5}
  \begin{ruledtabular}
    \begin{tabular}{l|cccc}
      \multirow{2}{*}{Mode} & $\varepsilon$ & $N^{}_{\rm sig}$ & $N^{\rm UL}_{\rm sig}$ & ${\cal B}^{\rm UL}$    \\
      & (\%) &  &  &  $(10^{-7})$ \\
      \hline
      $B^{0} \!\to\! K^{\ast 0}\mu^{+}e^{-}$ &      8.8  & $-1.5^{+4.7}_{-4.1}$ &  5.2 & 1.2 \\
      $B^{0} \!\to\!  K^{\ast 0}\mu^{-}e^{+}$ &       9.3 & $0.4^{+4.8}_{-4.5}$  & 7.4 & 1.6  \\
      $B^{0} \!\to\!  K^{\ast 0}\mu^{\pm} e^{\mp}$ (combined) & 9.0 & $-1.2^{+6.8}_{-6.2}$ & 8.0  & 1.8  \\
    \end{tabular}
  \end{ruledtabular}
\end{table}

There are a number of systematic uncertainties, as listed in Table~\ref{tab:syst}. The uncertainty on $\varepsilon$ due to limited MC statistics is $0.3\%$, and the uncertainty on the number of $B^{0}\bar{B^{0}}$ pairs is $1.4\%$. The systematic uncertainties related to detector performance are determined from dedicated studies of control samples; specifically, these samples are used to measure tracking and particle identification efficiencies of charged particles. The systematic uncertainty due to charged track reconstruction is $0.35\%$ per track. The uncertainty due to particle identification requirements is $2.8\%$. The uncertainty due to the requirements imposed on ${\cal O}^{q\bar{q}}_{\rm NN}$ and ${\cal O}^{\rm BB}_{\rm NN}$ is evaluated by imposing the same requirements on the control sample of $B\to K^{*0} J/\psi,\,J/\psi\to\ell^+\ell^-$ decays. We compare the efficiencies of the ${\cal O}^{}_{\rm NN}$ criteria on the control sample to those obtained from corresponding MC samples; the ratio is used to correct our signal efficiency, and the statistical error on the ratio is taken as the systematic uncertainty. For ${\cal O}^{q\bar{q}}_{\rm NN}$, this ratio is $1.002\,\pm 0.022$; for ${\cal O}^{\rm BB}_{\rm NN}$, the ratio is $0.919\,\pm 0.026$. The total systematic uncertainty due to both NN criteria applied together is 2.8\%. The uncertainty due to the PDF shapes is evaluated by varying the fixed PDF shape parameters by $\pm 1 \sigma$ and repeating the fit; the change in the central value of $N^{}_{\rm sig}$ is taken as the systematic uncertainty. Systematic uncertainties due to the aforementioned tiny contribution of the charmless hadronic $B$ decays are included. We initially assume that the $K^{\ast 0}$ is unpolarized. To investigate the effect of this, we calculate the reconstruction efficiency for fully longitudinal and fully transverse polarizations. The efficiency varies by only a few percent, and we include this variation as a systematic uncertainty.

Our reconstruction efficiency corresponds to  $B^{0}\to K^{\ast 0} \mu^{\pm} e^{\mp}$ decays proceeding according to three-body phase space. The corresponding $q^{2} \equiv M^{2}(\ell^{+}\ell^{-})$ spectra peak at low values, where the reconstruction efficiency is also low; thus our upper limits are conservative. For larger values of $q^{2}$, the efficiency rises approximately linearly from a minimum of 8\% to 14\% near $q^{2}_{\rm max}$. Such higher efficiencies would give lower upper limits. 

\begin{table}[tbh]
\caption{Systematic uncertainties included in calculating the upper limits. }
\label{tab:syst}
\renewcommand{\arraystretch}{1.3}
\begin{ruledtabular}
\begin{tabular}{l|ccc}
  \multirow{2}{*}{Source} &  \multicolumn{3}{c}{Systematic uncertainty (\%)} \\
  \cline{2-4}
  & $K^{\ast 0} \mu^{+} e^{-}$ & $K^{\ast 0}\mu^{-}e^{+}$ & $K^{\ast 0}\mu^\pm e^\mp$ \\
  \hline
  Reconstruction efficiency   & $\pm 0.3$ &  $\pm 0.3$ & $\pm 0.3$ \\
  Number of $B^{0}\bar{B^{0}}$ pairs  & $\pm 1.4$  &$\pm 1.4$ &$\pm 1.4$ \\
  $f^{00}$                          & $\pm 1.2$  &$\pm 1.2$ &$\pm 1.2$ \\
  Track reconstruction       & $\pm 1.4$  &$\pm 1.4$ &$\pm 1.4$ \\
  Particle identification    & $\pm 2.8$ & $\pm 2.8$ & $\pm 2.8$ \\
  ${\cal O}^{q\bar{q}}_{\rm NN}$ and ${\cal O}^{\rm BB}_{\rm NN}$   & $\pm 2.8$ & $\pm 2.8$ & $\pm 2.8$ \\
  PDF shape parameters        & $^{+2.1}_{-3.0}$ & $^{+8.2}_{-8.1}$ & $^{+4.5}_{-4.5}$\\
  $B \to$ charmless decays & $\pm 0.5$ & $\pm 2.2$ & $\pm 1.4$ \\
  $K^{\ast 0}$ polarization & $^{+2.7}_{-1.4}$ & $^{+3.8}_{-1.9}$ & $^{+3.2}_{-1.6}$ \\
  \hline
  Total  & $^{+5.7}_{-5.6}$ & $^{+10.3}_{-9.7}$ & $^{+7.2}_{-6.7}$\\      
\end{tabular}
\end{ruledtabular}
\end{table}

In summary, we have searched for the lepton-flavor-violating decays $B^{0}\to K^{\ast 0} \mu^{\pm} e^{\mp}$ using the full Belle data set recorded at the \Y4S resonance. We see no statistically significant signal and set the following 90\% C.L. upper limits on the branching fractions:
\begin{eqnarray}
{\cal B}(B^{0}\to K^{\ast 0} \mu^{+} e^{-}) & < & 1.2 \times 10^{-7} \\  
{\cal B}(B^{0}\to K^{\ast 0} \mu^{-} e^{+}) & < & 1.6 \times 10^{-7} \\  
{\cal B}(B^{0}\to K^{\ast 0} \mu^{\pm} e^{\mp}) & < & 1.8 \times 10^{-7}\,.
\end{eqnarray}
These results are the most stringent constraints on these LFV decays to date.

\section{Acknowledgments}

We thank the KEKB group for the excellent operation of the
accelerator; the KEK cryogenics group for the efficient
operation of the solenoid; and the KEK computer group,
the National Institute of Informatics, and the 
Pacific Northwest National Laboratory (PNNL) Environmental Molecular Sciences Laboratory (EMSL) computing group for valuable computing
and Science Information NETwork 5 (SINET5) network support.  We acknowledge support from
the Ministry of Education, Culture, Sports, Science, and
Technology (MEXT) of Japan, the Japan Society for the 
Promotion of Science (JSPS), and the Tau-Lepton Physics 
Research Center of Nagoya University; 
the Australian Research Council;
Austrian Science Fund under Grant No.~P 26794-N20;
the National Natural Science Foundation of China under Contracts
No.~11435013,  
No.~11475187,  
No.~11521505,  
No.~11575017,  
No.~11675166,  
No.~11705209;  
Key Research Program of Frontier Sciences, Chinese Academy of Sciences (CAS), Grant No.~QYZDJ-SSW-SLH011; 
the  CAS Center for Excellence in Particle Physics (CCEPP); 
Fudan University Grant No.~JIH5913023, No.~IDH5913011/003, 
No.~JIH5913024, No.~IDH5913011/002;                        
the Ministry of Education, Youth and Sports of the Czech
Republic under Contract No.~LTT17020;
the Carl Zeiss Foundation, the Deutsche Forschungsgemeinschaft, the
Excellence Cluster Universe, and the VolkswagenStiftung;
the Department of Science and Technology of India; 
the Istituto Nazionale di Fisica Nucleare of Italy; 
National Research Foundation (NRF) of Korea Grants No.~2014R1A2A2A01005286, No.2015R1A2A2A01003280,
No.~2015H1A2A1033649, No.~2016R1D1A1B01010135, No.~2016K1A3A7A09005 603, No.~2016R1D1A1B02012900; Radiation Science Research Institute, Foreign Large-size Research Facility Application Supporting project and the Global Science Experimental Data Hub Center of the Korea Institute of Science and Technology Information;
the Polish Ministry of Science and Higher Education and 
the National Science Center;
the Ministry of Education and Science of the Russian Federation and
the Russian Foundation for Basic Research;
the Slovenian Research Agency;
Ikerbasque, Basque Foundation for Science, Basque Government (No.~IT956-16) and
Ministry of Economy and Competitiveness (MINECO) (Juan de la Cierva), Spain;
the Swiss National Science Foundation; 
the Ministry of Education and the Ministry of Science and Technology of Taiwan;
and the United States Department of Energy and the National Science Foundation.


\begin{thebibliography}{99}
\bibitem{btosll:lhcb}
  R. Aaij \etal (LHCb Collaboration),
  Phys. Rev. Lett. {\bf 113}, 151601 (2014).

\bibitem{btosll:lhcb1}
  R. Aaij \etal (LHCb Collaboration),
  J. High Energy Phys. {\bf 08} (2017) 055.

\bibitem{btosll:th1}
  W. Altmannshofer, P. Ball, A. Bharucha, A. J. Buras, D. M. Straub, and M. Wick,
  J. High Energy Phys. {\bf 01} (2009) 019.

\bibitem{btosll:th2}
  W. Altmannshofer and D. M. Straub,
  Eur.\ Phys.\ J.\ C {\bf 75}, 382 (2015).

\bibitem{btosll:th3}
  S. Descotes-Genon, J. Matias, M. Ramon, and J. Virto,
  J. High Energy Phys. {\bf 01} (2013) 048.

\bibitem{btosll:th4}
  S. Descotes-Genon, T. Hurth, J. Matias, and J. Virto,
  J. High Energy Phys. {\bf 05} (2013) 137.

\bibitem{btosll:th5}
  S. Descotes-Genon, L. Hofer, J. Matias, and J. Virto,
  J. High Energy Phys. {\bf 12} (2014) 125.

\bibitem{btosll:th6}
  B. Capdevila, S. Descotes-Genon, J. Matias, and J. Virto,
  J. High Energy Phys. {\bf 10} (2016) 075.

\bibitem{btosll:th7}
  S. Descotes-Genon, L. Hofer, J. Matias, and J. Virto,
  J. High Energy Phys. {\bf 06} (2016) 092.

\bibitem{btosll:th8}
  B. Capdevila, A. Crivellin, S. Descotes-Genon, J, Matias, and Javier Virto,
  J. High Energy Phys. {\bf 01} (2018) 093.

\bibitem{lfv:th1}
  S. L. Glashow, D. Guadagnoli, and K. Lane,
  Phys.\ Rev.\ Lett.\ {\bf 114}, 091801 (2015).

\bibitem{lfv:th2}
  D. Guadagnoli and K. Lane,
  Phys.\ Lett.\ B {\bf 751}, 54 (2015).
  
\bibitem{lfv:th3}
  S. M. Boucenna, J. W. F. Valle, and A. Vicente,
  Phys.\ Lett.\ B {\bf 750}, 367 (2015).

\bibitem{lfv:th4}
  B. Gripaios, M. Nardecchia, and S. A. Renner,
  J. High Energy Phys. {\bf 05} (2015) 006. 

\bibitem{lfv:th5}
  B. Bhattacharya, A. Datta, D. London and S. Shivashankara
  Phys.\ Lett.\ B {\bf 742}, 370 (2015).

\bibitem{lfv:th6}
  S. Sahoo and R. Mohanta,
  Phys.\ Rev.\ D {\bf 91}, 094019 (2015).

\bibitem{lfv:th7}
  I. de Medeiros Varzielas and G. Hiller,
  J. High Energy Phys. {\bf 06} (2015) 072.

\bibitem{lfv:th8}
  D. Be{\v{c}}irevi{\'{c}}, O. Sumensari, and R. Z. Funchal,
   Eur.\ Phys.\ J.\ C {\bf 76}, 134 (2016).

\bibitem{lfv:th9}
  A. Crivellin, D. M{\"u}ller, A. Signer, and Y. Ulrich
  Phys.\ Rev.\ D {\bf 97}, 015019 (2018).

\bibitem{kst892}
  The $K^{\ast}(892)^{0}$ is denoted as $K^{\ast 0}$ throughout this paper.
  
\bibitem{lfv:babar}
  B. Aubert \etal (BaBar Collaboration),
  Phys.\ Rev.\ D {\bf 73}, 092001 (2006).

\bibitem{KEKB}
  S. Kurokawa and E. Kikutani, Nucl.\ Instrum.\ Methods Phys. Res., Sec. A {\bf 499}, 1 (2003), and other papers included in this volume; T. Abe \etal, Prog. Theor. Exp. Phys. {\bf 2013}, 03A001 (2013) and following articles up to 03A011.

\bibitem{belle:detector}
  A. Abashian \etal (Belle Collaboration), Nucl.\ Instrum.\ Methods Phys. Res., Sec. A {\bf 479}, 117 (2002); also, see the detector section in J. Brodzicka \etal, Prog. Theor. Exp. Phys. {\bf 2012}, 04D001 (2012).

\bibitem{nbb}
   Z. Natkaniec \etal (Belle SVD2 Group), Nucl.\ Instrum.\ Methods Phys. Res., Sec. A {\bf 560}, 1 (2006).
  
\bibitem{evtgen}
  D. J. Lange,
  Nucl.\ Instrum.\ Methods Phys. Res., Sec. A {\bf  462}, 152 (2001).

\bibitem{geant3}
  R. Brun \etal, CERN Report No. DD/EE/84-1 (1984).

\bibitem{cc}
  The inclusion of the charge conjugate decay mode is implied unless otherwise stated.

\bibitem{muid}
  A. Abashian \etal,
  Nucl.\ Instrum.\ Methods Phys. Res., Sec. A {\bf 491}, 69 (2002).

\bibitem{pid}
  E. Nakano,
  Nucl.\ Instrum.\ Methods Phys. Res., Sec. A {\bf 494}, 402 (2002). 

\bibitem{eid}
  K. Hanagaki, H. Kakuno, H. Ikeda, T. Iijima and T. Tsukamoto,
  Nucl.\ Instrum.\ Methods Phys. Res., Sec. A {\bf 485}, 490 (2002).

\bibitem{pdg}
  C. Patrignani \etal (Particle Data Group), Chin. Phys. C {\bf 40}, 100001 (2016) and 2017 update.

\bibitem{KSFW}
  S. H. Lee \etal (Belle Collaboration),
  Phys.\ Rev.\ Lett.\ {\bf 91}, 261801 (2003).

\bibitem{FW}
  G. C. Fox and S. Wolfram,
  Phys.\ Rev.\ Lett.\ {\bf 41}, 1581 (1978)

\bibitem{belle:qr}
  H. Kakuno \etal (Belle Collaboration),
  Nucl.\ Instrum.\ Methods Phys. Res., Sec. A {\bf 533}, 516 (2004).  

\bibitem{argus}
  H.~Albrecht \etal (ARGUS Collaboration),
  Phys.\ Lett.\ B {\bf 241} (1990) 278.
    
\end{thebibliography}
\end{document}